\documentclass[sn-mathphys-ay]{sn-jnl}

\usepackage{graphicx}%
\usepackage{multirow}%
\usepackage{amsmath,amssymb,amsfonts}%
\usepackage{amsthm}%
\usepackage{mathrsfs}%
\usepackage[title]{appendix}%
\usepackage{longtable}
\usepackage{multicol}
\usepackage[table]{xcolor}%
\usepackage{textcomp}%
\usepackage{manyfoot}%
\usepackage{booktabs}%
\usepackage{algorithm}%
\usepackage{algorithmicx}%
\usepackage{algpseudocode}%
\usepackage{listings}%
\usepackage{subcaption}

\begin{document}

\title[GNS for Aerosol Dynamics]{Learning to Simulate Aerosol Dynamics with Graph Neural Networks}

\author[1,2]{\fnm{Fabiana} \sur{Ferracina}}\email{fabiana.ferracina@wsu.edu}
\author[1]{\fnm{Payton} \sur{Beeler}}\email{payton.beeler@pnnl.gov}
\author[3]{\fnm{Mahantesh} \sur{Halappanavar}}\email{hala@pnnl.gov}
\author[2]{\fnm{Bala} \sur{Krishnamoorthy}}\email{kbala@wsu}
\author[3]{\fnm{Marco} \sur{Minutoli}}\email{marco.minutoli@pnnl.gov}
\author*[1]{\fnm{Laura} \sur{Fierce}}\email{laura.fierce@pnnl.gov}

\affil*[1]{\orgdiv{Atmospheric, Climate, and Earth Sciences Division}, \orgname{Pacific Northwest National Laboratory}, \orgaddress{\street{902 Battelle Boulevard}, \city{Richland}, \postcode{99352}, \state{Washington}, \country{USA}}}
\affil[2]{\orgdiv{Department of Mathematics and Statistics}, \orgname{Washington State University}, \orgaddress{\street{14204 NE Salmon Creek Ave}, \city{Vancouver}, \postcode{98686}, \state{Washington}, \country{USA}}}
\affil[3]{\orgdiv{Advanced Computing, Mathematics, and Data Division}, \orgname{Pacific Northwest National Laboratory}, \orgaddress{\street{902 Battelle Boulevard}, \city{Richland}, \postcode{99352}, \state{Washington}, \country{USA}}}

\abstract{Aerosol effects on climate, weather, and air quality depend on characteristics of individual particles, which are tremendously diverse and change in time. Particle-resolved models are the only models able to capture this diversity in particle physiochemical properties, and these models are computationally expensive. As a strategy for accelerating particle-resolved microphysics models, we introduce Graph-based Learning of Aerosol Dynamics (GLAD) and use this model to train a surrogate of the particle-resolved model PartMC-MOSAIC. GLAD implements a Graph Network-based Simulator (GNS), a machine learning framework that has been used to simulate particle-based fluid dynamics models. In GLAD, each particle is represented as a node in a graph, and the evolution of the particle population over time is simulated through learned message passing. We demonstrate our GNS approach on a simple aerosol system that includes condensation of sulfuric acid onto particles composed of sulfate, black carbon, organic carbon, and water. A graph with particles as nodes is constructed, and a graph neural network (GNN) is then trained using the model output from PartMC-MOSAIC. The trained GNN can then be used for simulating and predicting aerosol dynamics over time. Results demonstrate the framework's ability to accurately learn chemical dynamics and generalize across different scenarios, achieving efficient training and prediction times. We evaluate the performance across three scenarios, highlighting the framework's robustness and adaptability in modeling aerosol microphysics and chemistry.}

\maketitle

\section{Introduction}
\label{sec:introduction}
Aerosol particles, which are solid or liquid particles suspended in a gas, are a large source of uncertainty in understanding human impacts on climate \citep{masson2021climate,stier2013host}. Particles influence Earth's energy balance directly by scattering and absorbing solar radiation \citep{chung2005global,bellouin2005global} and indirectly by modifying the reflectivity and lifetime of clouds \citep{twomey1977influence,albrecht1989aerosols,rosenfeld2008flood}. A particle's optical properties and its ability to cloud droplets or ice crystals depends on its size, shape, and chemical composition. Atmospheric aerosol particles exhibit tremendous variability in these particle-level characteristics \citep[e.g.]{schwarz2008measurement,healy2014single,gautam2024chemical}. This particle-to-particle variability in physiochemical properties impacts climate-relevant aerosol properties, but is not easily represented in models \citep{riemer2019aerosol}. 

Particle-resolved models have emerged as a modeling strategy for simulating the evolution of complex particle distributions \citep[e.g.]{riemer2009simulating,shima2009super}. The Particle Monte Carlo Model for Simulating Aerosol Interactions and Chemistry (PartMC-MOSAIC) \citep{riemer2009simulating,zaveri2008development} tracks the size and composition of individual Monte Carlo particles as they evolve during transport. Similar Monte Carlo approaches have emerged for simulations of cloud particles, often implemented using the Super Droplet method introduced by \citet{shima2009super}.

Although particle-resolved models of aerosol and cloud particles have elucidated the role of small-scale processes on large-scale aerosol and cloud effects \citep[e.g.]{grabowski2017broadening,fierce2017toward,riemer2019aerosol,fierce2020radiative,morrison2020confronting,zheng2021quantifying,fierce2024quantifying}, these detailed models are computationally expensive. To accurately represent multivariate aerosol size-composition distributions, PartMC-MOSAIC must track the evolution of $10^4$--$10^6$ particles that interact with each other and with co-simulated trace gases. 

As a strategy to accelerate particle-resolved models, we introduce Graph-based Learning of Aerosol Dynamics (GLAD), a novel graph neural network (GNN) approach for learning aerosol dynamics on a per-particle level. Once the behavior is learned and represented on a graph, we can use this graph to simulate aerosol particle microphysics and chemistry. By encoding aerosol systems as graphs, with particles represented as nodes and interactions represented as edges, GNNs can capture spatial and temporal dependencies between particles and their interactions with the environment. 

The GLAD framework builds on previous approaches that apply GNNs and graph network simulators (GNS) to represent multiscale interactions. For example, GNNs have been used to simulate molecular dynamics, where they predict the behavior of particles over time by learning from the interactions within the system \cite{battaglia2016interaction}. In chemistry, GNNs have found applications in predicting molecular properties and reactions. By representing molecules as graphs, where atoms are nodes and bonds are edges, GNNs can learn to predict properties such as molecular energy, stability, and reactivity \cite{gilmer2017neural}. For example, GNNs have been utilized to predict the outcomes of chemical reactions, aiding in the design of new compounds with desired properties \cite{schutt2017quantum}.

The development of graph network simulators (GNS) for modeling complex physical systems offer new possibilities for simulating dynamic interactions in evolving systems. \cite{sanchez2020learning} introduced a framework that utilizes GNNs to simulate physical interactions, demonstrating the capability of GNNs to accurately model dynamics in particle-based systems by representing particles as nodes and interactions as edges. Building on this foundational work, \cite{kumar2022gns} developed the Generalizable Graph Neural Network-based Simulator, which extends the principles of GNNs to more generalizable and scalable simulations for particulate and fluid modeling. Their approach leverages the inherent flexibility of GNNs to handle different types of interactions and scales, making it particularly suitable for modeling complex systems like aerosols. 

The GLAD framework introduced here leverages GNS in a new way to accelerate particle-resolved simulations of aerosol dynamics. As a proof of concept, we focus on an idealized aerosol system consisting of sulfuric acid condensation onto black and organic particles and the subsequent water uptake. We hope to extend this approach to more complex particle systems in the future, which will make it possible to simulate more particles, to incorporate more complete representation of complex processes, and to perform the large simulation ensembles needed to train machine learning models for large-scale atmospheric models. 

\section{Approach}

Our approach leverages the capabilities of Graph Neural Networks to address the challenging and computationally expensive modeling of particle dynamics. By representing aerosol particles along with their chemical compositions (nodes) and their similarities and interactions (edges) as a graph, we aim to improve the computational efficiency of aerosol microphysics and chemistry simulations. 

\subsection{Simulating dynamics with GNNs}
\label{sub:introGNN}

GNNs have emerged as a powerful tool for modeling and analyzing data with inherent graph structures. They extend traditional neural networks by incorporating graph-based data, allowing for the effective learning of representations that capture relational information \cite{scarselli2008graph}. GNNs have been successfully applied across various domains, including social networks, biology, and physics, where data naturally form graph structures \cite{wu2020comprehensive}.

GNNs consist of multiple layers, each processing information from neighboring nodes and edges to update node, edge or global representations. Their architecture typically includes message passing mechanisms, aggregation functions, and learnable parameters. The neural networks can operate on both homogeneous and heterogeneous graphs, capturing complex relationships between entities in the graph structure \citep{wu2020comprehensive}. By iteratively aggregating information from neighboring nodes and edges, GNNs can learn representations that encode the structural and semantic properties of the graph. Training involves optimizing the parameters of the network to minimize a predefined loss function, with objectives ranging from node classification to graph-level prediction tasks. Recent advancements in GNN architectures, such as message passing neural networks and attention mechanisms, have further enhanced their capability to model complex graph-based data \cite{velivckovic2017graph}. These innovations have improved the accuracy and scalability of GNNs, making them a valuable tool in both microphysics and chemistry research.

Simulators built with neural networks, particularly Graph Neural Networks (GNNs), have shown significant promise in various domains due to their ability to model complex systems with relational data structures. The operation of a graph network simulator involves training a graph neural network (GNN) on historical data to learn the underlying dynamics of the system. During training, the parameters of the neural network are optimized using gradient-based algorithms to minimize a specified loss function, which measures the disparity between the predicted and actual values of the target variables \citep{wu2020comprehensive}. Once trained, the GNN can simulate the dynamic evolution of the system over time by iteratively updating the states of the nodes and edges based on the learned relationships encoded in its parameters. This iterative updating allows the model to capture and predict complex temporal changes within the system \citep{sanchez2020learning}.

\subsection{Procedure for training and using GNS of aerosol dynamics}
\label{sec:gnsframe}
The GNS framework uses graph neural networks for learning to simulate the chemical composition changes going on within the particles. Our GNN uses an Encode-Process-Decode architecture with a step to pre-process the data before it is fed into the encoder and a step to post-process the data after it is output by the decoder.

\subsubsection{Pre-processing}
\label{sub:preprocessing}
Pre-processing involves computing the rate of change in mass or concentration for each chemical in the particle or gas in the system, over two time steps, that is $\Delta m^C = (m^C_{t} - m^C_{t-1})/(t - (t-1))$. These $\Delta m^C$'s along particle properties $p$, which are properties that differ by particle but do not vary with time, as well as masses (or concentrations) differences from the known minima and maxima, are combined in the node data of our GNN. This means that each node is a particle who holds information about its own specific properties, and of its mass changes over time. 

We connect our network nodes using their Euclidean distance to two closest neighbors in the chemical composition space. The chemical composition space refers to the multi-dimensional space where each dimension represents the concentration or amount of a particular element or compound within a mixture or material. Although, not usually an Euclidean space, through some care we may consider the composition space as locally Euclidean:
\begin{itemize}
    \item[-] When considering small variations in composition, the space can be approximated as Euclidean. For instance, in a quaternary system (four elements), the composition space is a tetrahedron. Within a small region of this tetrahedron, changes in composition can be linearly interpolated, and distances can be measured using standard Euclidean metrics.
    \item[-] For a system with $n$ components, the composition can be represented in an $(n-1)$-dimensional simplex. For example, for a ternary system, this is a 2-dimensional triangle, and for a quaternary system, this is a 3-dimensional tetrahedron. Within small regions of these simplices, the space can be treated as locally Euclidean.
    \item[-]  If the chemical composition space is embedded in a metric space where distances between points (compositions) are defined in a way that obeys Euclidean properties (e.g., using the Euclidean distance formula), then locally, the space can behave like a Euclidean space.
\end{itemize}

In the pre-processing step, we also combine data that will serve as edge representation. Specifically, we store in the edge structure information on the displacement and $\ell_2$ distances between the two nodes which the edge connects.

\subsubsection{Encode-Process-Decode Architecture}
\label{sub:encodeprocessdecode}

Once the pre-processing is done, we can move to the Encode step. The Encoder is an embedding function that takes the chemistry-state representations on the nodes and edges, $X$ and assigns them to a latent graph $G^0$. We denote the graph as $G^m = (V, E)$ with vertices $\mathbf{v}_i = f_v(\mathbf{x}_i) \in V$ and edges $\mathbf{e}_{i,j} = f_e(\mathbf{d}_{i,j}) \in E$. The functions $f_v$ and $f_e$ are learned functions that embed the particle chemistry information $(\mathbf{x}_i, \mathbf{d}_{i,j})$ into the latent nodes and edges respectively. 

The Process phase is when interactions amongst the connected nodes occur through a sequence of $M$ message-passing steps. In this phase, the Processor takes $G^0$ as its input, computes $(G^1, \dots, G^M)$, and outputs the final $G^M$. In each graph of the sequence, messages (rules and constraints) are propagated, information is aggregated, and nodes and edges are updated. The interaction network starts with learning the message embeddings on the nodes. In general, for node $i$, we have:
\begin{equation}
    \mathbf{x}_i^m = \xi^m \left(\mathbf{x}_i^{m-1}, \bigoplus\limits_{j \in \mathcal{N}(i)} \phi^m \left(\mathbf{x}_i^{m-1}, \mathbf{x}_j^{m-1}, \mathbf{e}_{j,i}\right) \right),
\end{equation}
where the features $\mathbf{x}_i^m$ are updated through a differentiable function $\xi^m$ (e.g. a multi-layer perceptron as in our case). The updates takes the features for node $i$ from graph $m-1$ and combines them with new information using some sort of aggregator such as sum. For all other nodes $j$ in $i$'s neighborhood, apply a differentiable function $\phi^m$ which specifies how the information in node $i$ will combine with those of nodes $j$ and the respective edges $ \mathbf{e}_{j,i}$'s that connect them.

Once all message passing steps have been performed, the output graph $G^M$ is given to the Decoder. The Decode phase involves converting the latent graph representation $G^M$ back into the original space of particle properties. Note that we do not decode the edges as we wish to learn only node representations given our framework. The node contains the chemistry information, while the edges are simply tracking the particles proximity in chemical composition space. The Decoder computes:
\[
\hat{\mathbf{y}}_i = \delta^v(\mathbf{v}_i^M),
\]
where $\delta^v$ is a learned function that maps the latent node representation back to the predicted property $\hat{\mathbf{y}}_i$. In our framework, the dynamics extracted from the nodes, pertain to how fast the changes in masses or concentrations are occurring within each particle. After decoding, we compute the loss between the predicted rates and the target rates - target rates are computed using an Inverse Decoder from the ground-truth masses and concentrations. After training  for several steps to minimize the loss, the post-processing of the predicted rates involves Euler integration to obtain the predicted masses and concentrations. 

\begin{figure}
    \centering
    \includegraphics[width=\textwidth]{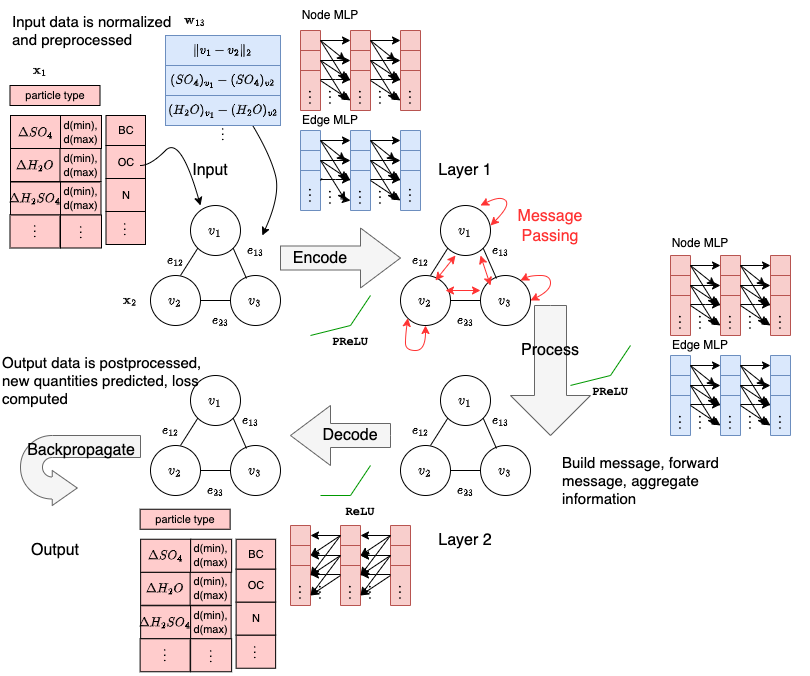}
    \caption{Architecture of a Graph Neural Network (GNN) for aerosol microphysics simulation. The process begins with input data normalization and preprocessing, followed by encoding using Multi-Layer Perceptrons (MLPs) for nodes and edges. Message passing is performed to exchange information between nodes, which is processed through multiple layers with activation functions like PReLU. The data is then decoded back into the original feature space, post-processed to predict new quantities, and refined through backpropagation to optimize model accuracy.}
    \label{fig:gnnsch}
\end{figure}

Figure \ref{fig:gnnsch} illustrates the architecture of a Graph Neural Network (GNN) designed for simulating aerosol microphysics. The process involves several key stages: input preprocessing, encoding, message passing, processing, decoding, and output postprocessing. The input data, consisting of various aerosol particles and their properties (e.g., $\Delta SO_4$, $\Delta H_2O$, $\Delta H_2SO_4$ - the rates of change in mass and concentration of each chemical species), are normalized and preprocessed. Each particle is further characterized by parameters such as type, chemical species' mass (or concentration) distances to minimum and maximum values ($d_{\text{min}}$, $d_{\text{max}}$), black carbon mass (BC), organic carbon mass (OC), and aerosol number concentration (N).

 The system constructs a graph where each node ($v_1, v_2, v_3$) represents a particle or a state, and edges ($e_{12}, e_{13}, e_{23}$) - added to the graph by a $k$-nearest neighbors algorithm - represent the Euclidean distance (i.e. $||v_i - v_j ||_2$) and displacement between all mass and concentration (e.g. $(SO_4)_{v_i} - (SO_4)_{v_j}$) quantities represented by the nodes. These input features are encoded using Multi-Layer Perceptrons (MLPs) for both nodes and edges, transforming them into a format suitable for processing within the GNN.

In the message passing phase, information is exchanged between nodes through edges. This step involves building messages, forwarding them along the edges, and aggregating information at each node. The process is iteratively applied across layers to capture complex interactions. The encoded data undergoes processing in Layer 1, where node and edge features are updated using MLPs. Activation functions such as Parametric ReLU (PReLU) are applied to introduce non-linearity.

The processed information is then decoded back into the original feature space using MLPs. This step translates the learned representations into predictions for the aerosol properties. The output data is post-processed to predict new quantities and compute the loss for model training. The predicted properties are compared to the target values to refine the model through backpropagation, which involves adjusting the weights of the network based on the computed loss, optimizing the model to improve accuracy in predicting aerosol microphysical properties.

\subsubsection{Simulating Aerosol Dynamics with a Learned Graph Network}
\begin{figure}
    \centering
    \includegraphics[width=\textwidth]{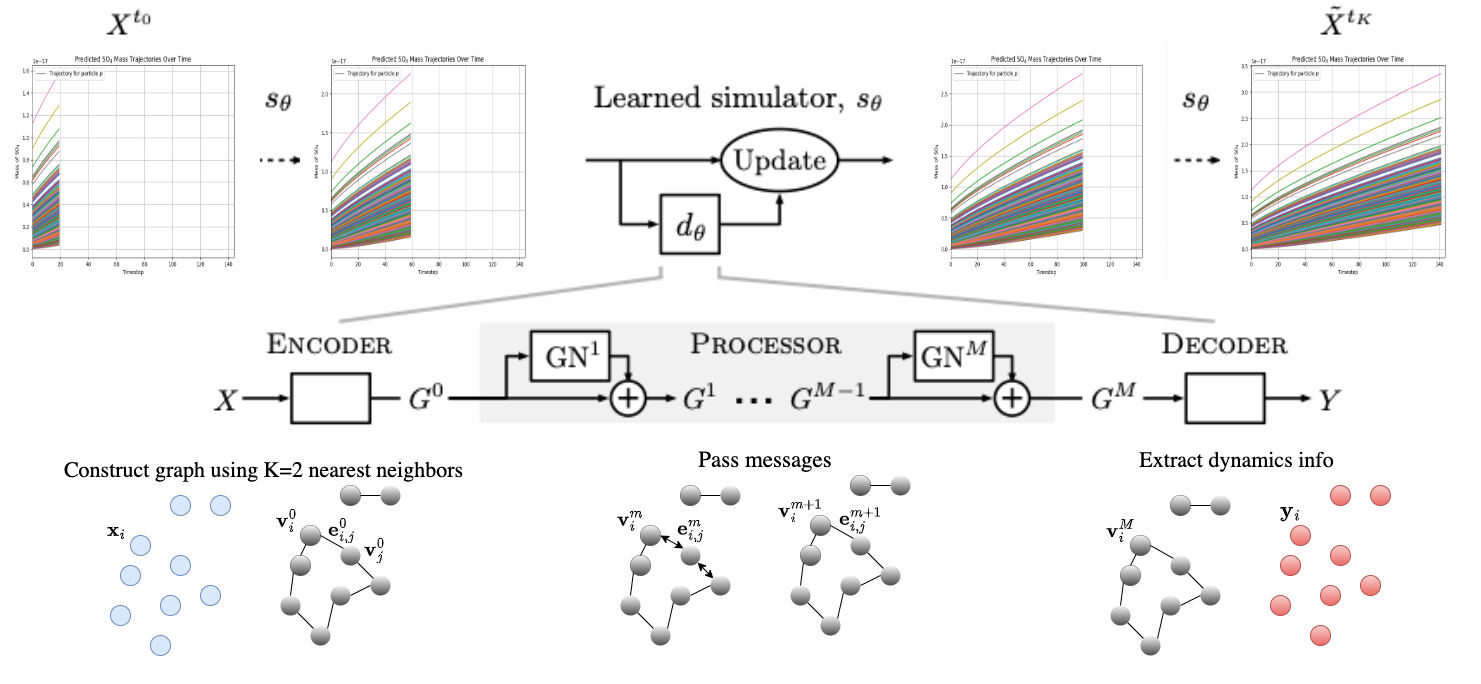}
    \caption{Schematics of our graph network simulator. This figure was adapted from \cite{sanchez2020learning} to depict our GNS framework. The first row shows the progress of the simulated data through time. At each time step the learned simulator is called to predict the next time step. The learned simulator is a pre-trained GNN using the Encode-Process-Decode scheme as shown in Figure \ref{fig:gnnsch}.}
    \label{fig:gnssch}
\end{figure}

Once the GNN is trained, it is then used as a simulator to predict aerosol microphysical dynamics iteratively. Figure \ref{fig:gnssch} illustrates the integration of a Graph Neural Network (GNN) within a graph network simulator (GNS) designed to predict aerosol microphysics over time. The GNN acts as the core computational engine, enabling the simulation of complex interactions between aerosol particles.

The simulator employs a learned update mechanism, denoted as $d_\theta$, to iteratively refine predictions. This mechanism adjusts the state of the system based on the GNN's output, ensuring accurate simulation of aerosol dynamics over time. The final output provides predictions of mass and concentration trajectories for various chemical species present in the aerosol particles, which are visualized over time. The simulator incorporates a feedback loop, using these predictions to update the graph and improve subsequent iterations.

\subsection{Target application: GNS surrogate of PartMC-MOSAIC}

We explore our GNS framework for aerosol dynamics with using data simulated by PartMC-MOSAIC. We explore a simple system where H$_2$SO$_4$ condenses on particles containing SO$_4$, black carbon (BC), organic carbon (OC) and H$_2$O. Within PartMC-MOSAIC, we simulate multiple 24-hour scenarios in which the initial concentrations of H$_2$SO$_4$ and total number concentration of particles were varied among scenarios. Data was recorded every 600 seconds, giving us a total of 144 time steps in which the gas interacts with the particles. The environmental properties are kept constant with relative humidity set to 95\%, temperature set to 293.15~K, and pressure set to 101325~Pa.

For each scenario, we train the GNS using the per-particle mass of OC, BC, SO$_4$, and H$_2$O, the number concentration associated with each computational particle, and the concentration of H$_2$SO$_4$. In these condensation-only scenarios, the particle number concentration, OC mass, and BC mass do not vary over time and, in our framework, are treated as ``material properties" of the particles. On the other hand, H$_2$SO$_4$ varies with time but does not vary among particles, such that each particle ``sees'' the same concentration of H$_2$SO$_4$ at a time.

Data from PartMC-MOSAIC is processed through our data preparation pipeline. Once the information has been loaded, each array is determined to be either a ``material property'' - a feature that does not change with time, but differs per particle, or a time changing feature which may or may not vary per particle. We can also specify ``particle types'' which can be used to encode a categorical characteristic for a particle such as CCN active or not, for instance, though this feature is not applied the examples presented here. For each scenario simulated in PartMC-MOSAIC, we assign a categorical variable (called ``universe number'') which is used by the algorithm to learn the differences between the scenarios.

Before taking the simulation data into our framework, we perform a data transformation step where gas concentrations are placed on a logarithmic scale and all quantities are normalized to fall in the $[0,1]$ interval. We also split the dataset into training, testing and validation scenarios. The splits can be determined as desired, but for the examples shown in our experiments we use a 60/30/10 train/test/validation split.

\subsection{Implementation Details}
\label{sub:imp}

As in the work of \cite{sanchez2020learning} and \cite{kumar2022gns}, our GNS uses standard neural networks and $k$-nearest neighbors ($k$-NN) algorithms as building blocks. We use PyTorch \citep{paszke2019pytorch} and PyTorch Geometric \citep{fey2019fast} classes and functions for these purposes.

As aforementioned, particle's input are a sequence of per-particle masses of constituent aerosol species or gas concentrations, which are used to compute the rate of change in these quantities over each time step. We also include particle features that do not change over time, so called ``material properties''. Unlike previous GNS implementations, our GNS framework is designed to operate in beyond a three-dimensional space. In our case, both the time-changing and material property features can be arrays of arbitrary dimensions. We also add a novel global categorical variable called ``universe number'', which represents the particles' scenario conditions.

Our Encoder constructs a $k$-NN graph with $k=2$, that is each particle is connected to the nearest 2 neighbors in the chemical composition domain, with necessary transformations applied so that we could use Euclidean distances. The data contained in nodes and edges are encoded as 2-hidden-layer perceptrons (MLP) with parametric rectified linear unit functions $\mathrm{PReLU}(x; a) = \max(0,x) + a \ast \min(0, x)$ applied at each of 256 channels, that learns the embeddings. 

The Processor performs $M=1$ message passing step, which means that the interaction network contains one message exchange between each of the nodes and their neighbors (and the edge that connects them). Both $k$ and $M$ are parameters on which we experimented for performance improvements. Messages are combined and exchanged with the same MLP from the Encoder, and are aggregated via addition.

Our Decoder is an alternative 2-hidden-layer perceptron whose activation functions are the regular non-parametric ReLU, that is $\delta^v(x) = \max(0, x)$. After being decoded, the data represent the dynamics of chemical compositions of each particle. The predicted compositions are obtained using an Euler integrator.

As in \cite{sanchez2020learning}'s work, we implement training noise to mitigate error accumulation over the rollouts. The ``ground-truth'' input data, which is PartMC-MOSAIC simulated data is not corrupted by noise. However, as our model predicts each time step based on its previous predictions, substantial error accumulation can occur. Mitigating this accumulation involves introducing noise to the training data. After computing the rates of changes, we add a random-walk noise that is distributed normally as $\mathcal{N}(\mu = 0, \sigma = 6.3 \times 10^{-5} )$, thus making the training distribution closer to the distribution generated from rollouts.

We also standardize all input and target vectors with means and standard deviations summarized from the training dataset. Without this transformation, we noticed poor performance due to the large mean and variability ranges of the multi-dimensional chemical composition space.

Loss at training was computed by comparing the predicted per-particle dynamics with target ones obtained from the ground-truth data. We use a multi-output mean squared error function to compute this loss:
\begin{equation}\label{eq:lmse}
\mathcal{L}_{\text{MSE}} = \frac{1}{N} \sum_{i=1}^{N} \sum_{j=1}^{D} \left( y_{ij} - \hat{y}_{ij} \right)^2,
\end{equation}
where $N$ is the total number of particles, $D$ is the total number of dimensions (i.e. chemistry in the particles), $y_{ij}$ is the target rate of chemical change based on ground-truth data, and $\hat{y}_{ij}$ is the predicted rate of chemical change.

We minimize this loss using the Adam optimization algorithm. Adam, which stands for Adaptive Moment Estimation, computes individual adaptive learning rates for different parameters by estimating the first moment (mean) and the second moment (uncentered variance) of the gradients \citep{kingma2014adam}. We set an initial learning rate of $5 \times 10^{-5}$ and perform a total of 6300 gradient update steps, 300 steps at a time, for each available scenario. 

\section{Results and Discussion}
\label{sec:results}
Our main findings are that a physics-based GNS can be adapted to work in multi-dimensional chemical composition space, and it can learn accurate chemistry dynamics information from just the initial values. It is generalizable to different scenarios and conditions for the same simple system of sulfuric acid condensation. Although the system used here is simple, so is our GNS, which does not specify any formulas or constraints on how the particles should behave.

Our GNS is also efficient. Training on GPU usually took less than 12 seconds when using three aggregated scenarios with a total of 7267 particles and performing 200 training steps, while testing and prediction time was always under 1 second for all the results presented here. The GNS framework is also quite modular, allowing for many of the parameters and hyperparameters to be set as appropriate or tuned differently. This allowed us to quickly experiment with different inputs to obtain the better performance.

\subsection{Demonstrating performance with three example scenarios}
The results presented here are for three testing scenarios which were predicted using a model trained on three aggregated training scenarios. Each scenario has different initial conditions, and the trained model has no apriori knowledge of the testing scenarios' initial conditions. Although neural networks work best on predicting scenarios whose initial conditions have already been learned, we wanted to test our model's performance on examples spawned from completely novel conditions. Scenario input parameters are summarized in Table~\ref{tab:inputs}, and the performance measures for each scenario are provided in Table \ref{tab:scenumbers}. Accuracy of per-particle properties is quantified as the normalized mean absolute error (NMAE), defined as:
\begin{equation}
    \text{NMAE} = \frac{1}{m} \sum_{j=1}^{m} \frac{\sum_{i=1}^{n} \left| y_{ij} - \hat{y}_{ij} \right|}{\sum_{i=1}^{n} \left| y_{ij} \right|},
\end{equation}
where $\mathbf{y} = y_{ij}$ and $\hat{\mathbf{y}} = \hat{y}_{ij}$ are the true and predicted vectors, respectively, each of size $n$, the number of time steps by $m$, the number of particles. 

\begin{table}
    \centering
    \begin{tabular}{@{}lcc@{}}
    \hline
        Scenarios & Total \# particles & Initial H$_2$SO$_4$ Concentration (ppb) \\
    \hline
        Training Scenario 1 & 1975 & 2.3153 \\
        Training Scenario 2 & 2315 & 1.9610 \\
        Training Scenario 3 & 2977 & 4.3907 \\
        Testing Scenario 1 & 2019 & 5.7245 \\
        Testing Scenario 2 & 2829 & 9.2196 \\
        Testing Scenario 3 & 3821 & 9.9237 \\
    \hline
    \end{tabular}
    \caption{Total number of particles and initial H$_2$SO$_4$ concentrations in each of the scenarios presented here.}
    \label{tab:inputs}
\end{table}

\begin{table}
    \centering
    \begin{tabular}{@{}lcccc@{}}
    \hline
        Scenarios & \# particles & $\mathcal{L}_{\text{MSE}}$ & SO$_4$ NMAE & H$_2$SO$_4$ NMAE \\
        \hline
         Testing Scenario 1 & 2019& 0.0495& 0.1347 & 0.2660 \\ 
         Testing Scenario 2 & 2829&0.0211&0.0494& 0.3632\\ 
         Testing Scenario 3 & 3821 & 0.0003 & 0.1285 & 0.2270\\ 
         \hline
    \end{tabular}
    \caption{Results for a subset three of the nine scenarios tested. NMAE is the normalized mean absolute error between the prediction and the ground-truth data (PartMC-MOSAIC simulated data). SO$_4$'s NMAE is between masses in kg, H$_2$SO$_4$'s NMAE is between concentrations in ppb.}
    \label{tab:scenumbers}
\end{table}

\subsection{Per-particle composition and trace gases over time}
\begin{figure}[t]
    \centering
    \includegraphics[width=\textwidth]{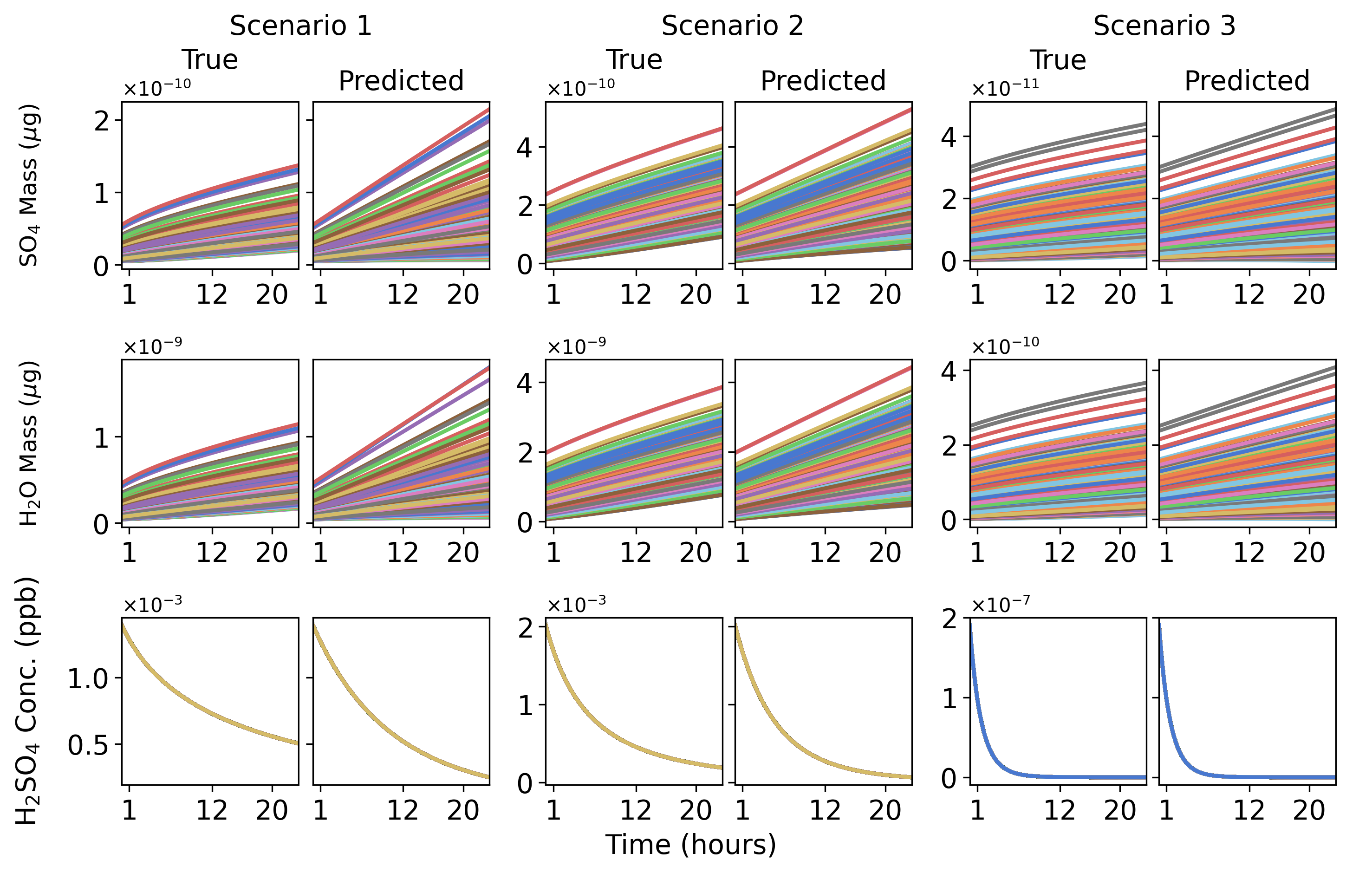}
    \caption{Comparison between ``ground-truth'' data (PartMC-MOSAIC accepted simulation results) and our GNS simulation for each scenario's SO$_4$ and H$_2$O masses in $\mu$g, and H$_2$SO$_4$ concentration in ppb. Each line represents the chemical mass/concentration trajectory for each particle over time for 24 hours. Trajectories where predicted using the first two masses/concentrations of each chemical for each particle.}
    \label{fig:all_sc_traj}
\end{figure}

As illustrated by Figure \ref{fig:all_sc_traj}, the GLAD model accurately predicts per-particle SO$_4$ and H$_2$O, as well as the corresponding concentration in H$_2$SO$_4$ in the first 4 to 6 hours - this is impressive given its small initial condition corpus of knowledge. In both the particle-resolved and GLAD model, the mass of SO$_4$ in individual particles as H$2$SO$_4$ partitions to the particle phase, corresponding to a decrease in the concentration of H$_2$SO$_4$ gas. The increase in per-particle SO$_4$ mass leads to an increase in the mass of water taken up by particles, again in both the PartMC-MOSAIC simulation and in the learned GNS model. 

While the model performs well overall in all scenarios, slight discrepancies at the extremes of the size range highlight areas for potential improvement. The SO$_4$ mass fraction ranges predominantly between 0.4 and 1.0 in all three scenarios, providing insight into the typical composition of these aerosol particles under the studied conditions. These observations underscore the complex interplay between particle size, composition, and atmospheric processing, while also demonstrating the efficacy of our predictive model in capturing these intricate dynamics.

\begin{figure}
    \centering
    \includegraphics[width=0.8\linewidth]{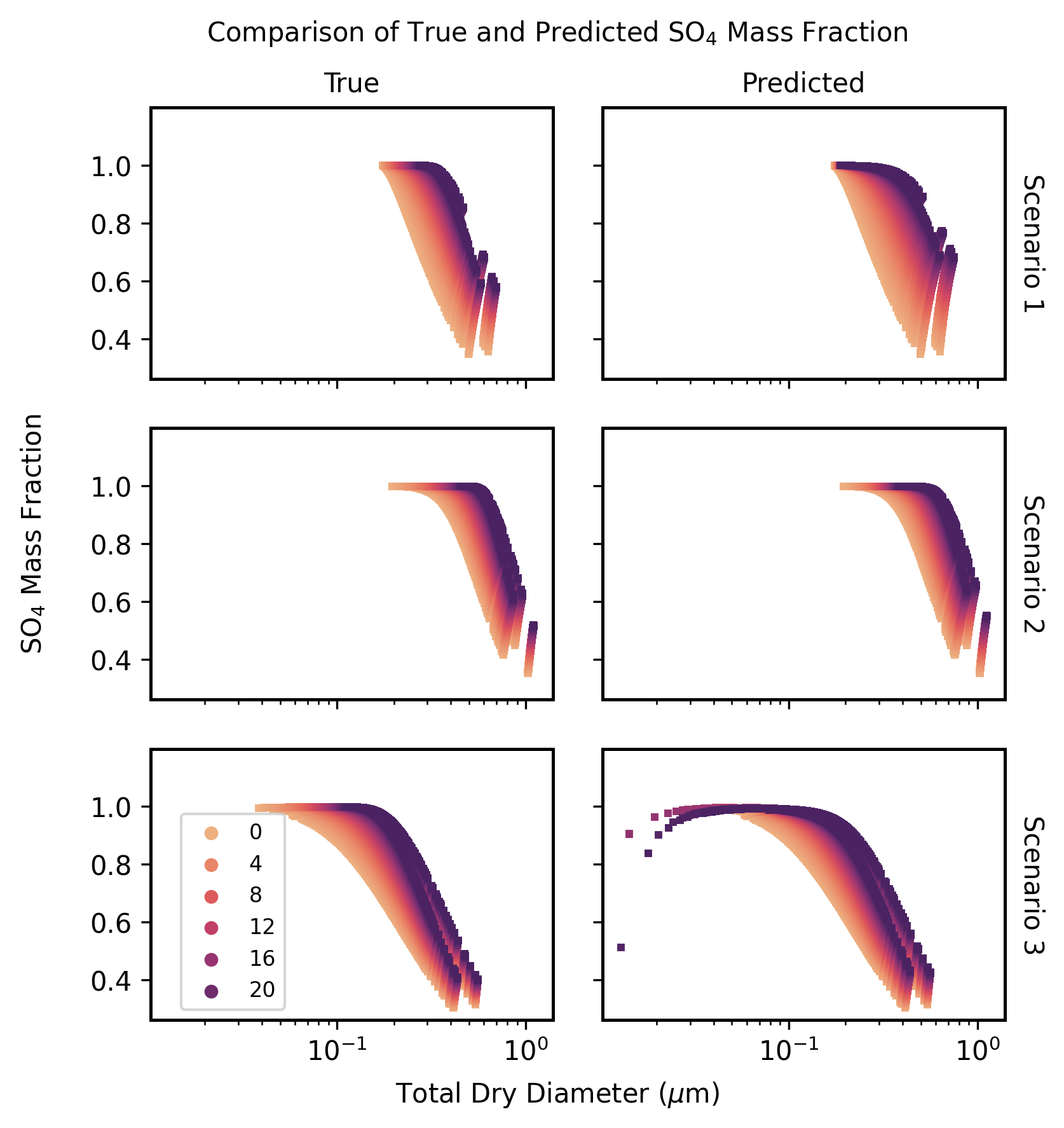}
    \caption{Comparison of true and predicted SO$_4$ mass fraction versus total dry diameter for aerosol particles over time for all three scenarios. The left panels show the true values for each scenario, while the right panels displays their respective predicted values. Each color represents a different time point ranging from 0 to 24 hours. The $x$-axis shows the total dry diameter in $\mu$m on a logarithmic scale, and the $y$-axis represents the SO$_4$ mass fraction.}
    \label{fig:scABC}
\end{figure}

\subsection{Evolution of the aerosol size distribution}
\begin{figure}
    \centering
    \includegraphics[width=0.9\linewidth]{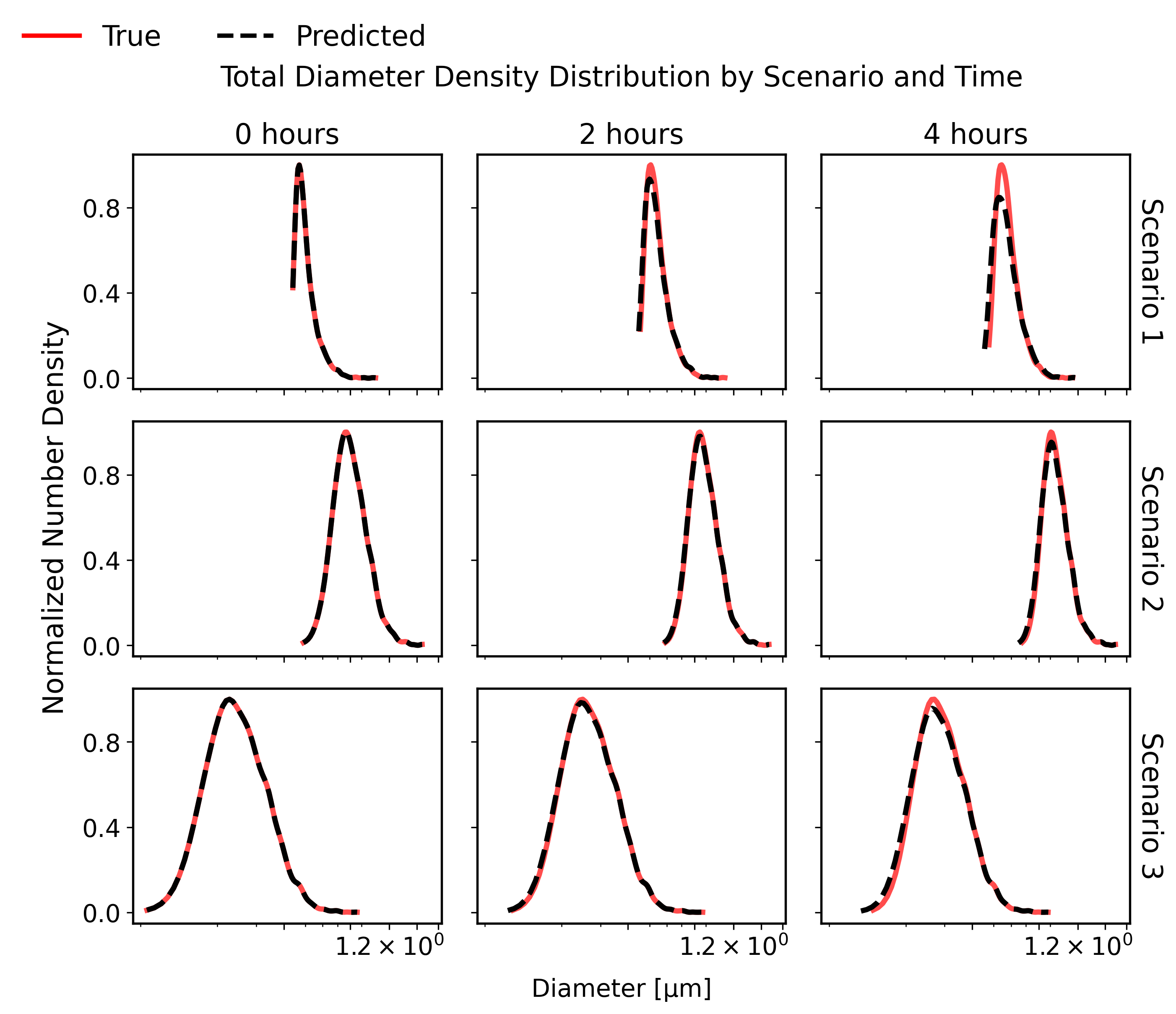}
    \caption{Comparison of true and predicted total diameter density distributions across three scenarios at different time intervals (0 hr, 2 hrs, and 4 hrs). Solid lines represent true distributions, while dashed lines indicate predictions by the simulator. Each row corresponds to a different scenario, demonstrating the model's ability to accurately capture aerosol dynamics over time.}
    \label{fig:diam_densities}
\end{figure}

With such accurate predictions in per-particle properties during early dynamics, it is unsurprising that the GLAD framework predicts the initial evolution of the aerosol size distribution nearly perfect accuracy in comparison with PartMC-MOSAIC (see Figure~\ref{fig:diam_densities}). The size distribution at different points in time (0 hr, 2 hrs, and 4 hrs) are shown for the three example scenarios, shown for the true PartMC-MOSAIC simulations (solid red lines) and the GLAD predictions (dashed black lines).

\section{Conclusion}
\label{sec:conclusion}

This study demonstrates the effectiveness of a physics-based Graph Neural Network (GNN) approach in modeling multi-dimensional chemical composition dynamics in aerosols. By adapting the Graph Network Simulator (GNS) framework to work with chemical composition space, we have shown that accurate chemistry dynamics information can be learned from initial conditions alone, without specifying explicit formulas or constraints on particle behavior.

Our GNS framework, applied to a simple system of sulfuric acid condensation, exhibits several key strengths:

\begin{itemize}
    \item[-] \emph{Accuracy and efficiency:} The model demonstrates relatively high accuracy across different scenarios it has never seen before. Although predictions diverge from truth over time, this divergence can be made smaller as the model's knowledge corpus grows. Training times are notably efficient, typically requiring approximately 12 seconds for 200 training steps to learn the chemistry dynamics of 7267 particles over a 24 hours period (on GPU), while testing and prediction times consistently remain under 1 second.
    
    \item[-] \emph{Generalizability:} The framework shows robust performance across various scenarios with different initial concentrations of H$_2$SO$_4$ and total number of particles, indicating its ability to adapt to diverse environmental conditions within the studied system.
    
    \item[-] \emph{Flexibility:} The modular nature of the GNS allows for easy adjustment of parameters and hyperparameters, facilitating rapid experimentation and optimization for different input conditions.
\end{itemize}

The comparison between true and predicted SO$_4$ mass fractions across different scenarios demonstrates the model's ability to capture complex relationships between particle size, composition, and temporal evolution. The striking similarities between true and predicted plots, particularly in representing the inverse relationship between particle size and SO$_4$ mass fraction, underscore the model's efficacy in simulating intricate aerosol dynamics.

As for future work, several promising directions emerge, such as

\begin{itemize}
    \item[-] Incorporating global nodes could provide a more efficient representation of uniform properties like H$_2$SO$_4$ concentration, potentially improving the model's performance and computational efficiency.
    
    \item[-] Exploring distance functions beyond Euclidean, such as Mahalanobis or cosine distances, may offer better representations of the complex relationships in chemical composition space, potentially leading to improved model performance and deeper insights into underlying chemical processes.
    
    \item[-] Extending the framework to more complex aerosol systems with additional chemical species and processes would test the limits of the GNS approach and potentially broaden its applicability in atmospheric microphysics and chemistry modeling.
\end{itemize}

Further investigation into optimal parameter settings could also yield improvements in prediction accuracy, particularly for scenarios that currently show higher prediction errors. In conclusion, this study presents a significant step forward in the application of machine learning techniques to aerosol microphysics and chemistry modeling. The demonstrated ability of GLAD - our GNS framework  - to accurately and efficiently simulate aerosol dynamics opens up new possibilities for understanding and predicting atmospheric processes. As we continue to refine and expand this approach, we anticipate that it will become an increasingly valuable tool in the study of atmospheric chemistry and climate science, offering new insights into the complex interactions that shape our environment.

\subsection*{Data availability}
The GLAD package is available at \url{github.com/fabstat/glad}. The repository also includes the PartMC-MOSAIC data that was used for training, validation, and testing, as well as the example test cases and plotting scripts. PartMC is available at \url{https://github.com/compdyn/partmc}, and MOSAIC is available from Rahul Zaveri (rahul.zaveri@pnnl.gov).

\subsection*{Acknowledgements}
The study was supported through the Laboratory Directed Research and Development program of Pacific Northwest National Laboratory (PNNL). PNNL is operated for the DOE by the Battelle Memorial Institute under Contract DE-AC05-76RL01830.

\bibliography{references.bib}


\begin{thebibliography}{31}
\ifx \bisbn   \undefined \def \bisbn  #1{ISBN #1}\fi
\ifx \binits  \undefined \def \binits#1{#1}\fi
\ifx \bauthor  \undefined \def \bauthor#1{#1}\fi
\ifx \batitle  \undefined \def \batitle#1{#1}\fi
\ifx \bjtitle  \undefined \def \bjtitle#1{#1}\fi
\ifx \bvolume  \undefined \def \bvolume#1{\textbf{#1}}\fi
\ifx \byear  \undefined \def \byear#1{#1}\fi
\ifx \bissue  \undefined \def \bissue#1{#1}\fi
\ifx \bfpage  \undefined \def \bfpage#1{#1}\fi
\ifx \blpage  \undefined \def \blpage #1{#1}\fi
\ifx \burl  \undefined \def \burl#1{\textsf{#1}}\fi
\ifx \doiurl  \undefined \def \doiurl#1{\url{https://doi.org/#1}}\fi
\ifx \betal  \undefined \def \betal{\textit{et al.}}\fi
\ifx \binstitute  \undefined \def \binstitute#1{#1}\fi
\ifx \binstitutionaled  \undefined \def \binstitutionaled#1{#1}\fi
\ifx \bctitle  \undefined \def \bctitle#1{#1}\fi
\ifx \beditor  \undefined \def \beditor#1{#1}\fi
\ifx \bpublisher  \undefined \def \bpublisher#1{#1}\fi
\ifx \bbtitle  \undefined \def \bbtitle#1{#1}\fi
\ifx \bedition  \undefined \def \bedition#1{#1}\fi
\ifx \bseriesno  \undefined \def \bseriesno#1{#1}\fi
\ifx \blocation  \undefined \def \blocation#1{#1}\fi
\ifx \bsertitle  \undefined \def \bsertitle#1{#1}\fi
\ifx \bsnm \undefined \def \bsnm#1{#1}\fi
\ifx \bsuffix \undefined \def \bsuffix#1{#1}\fi
\ifx \bparticle \undefined \def \bparticle#1{#1}\fi
\ifx \barticle \undefined \def \barticle#1{#1}\fi
\bibcommenthead
\ifx \bconfdate \undefined \def \bconfdate #1{#1}\fi
\ifx \botherref \undefined \def \botherref #1{#1}\fi
\ifx \url \undefined \def \url#1{\textsf{#1}}\fi
\ifx \bchapter \undefined \def \bchapter#1{#1}\fi
\ifx \bbook \undefined \def \bbook#1{#1}\fi
\ifx \bcomment \undefined \def \bcomment#1{#1}\fi
\ifx \oauthor \undefined \def \oauthor#1{#1}\fi
\ifx \citeauthoryear \undefined \def \citeauthoryear#1{#1}\fi
\ifx \endbibitem  \undefined \def \endbibitem {}\fi
\ifx \bconflocation  \undefined \def \bconflocation#1{#1}\fi
\ifx \arxivurl  \undefined \def \arxivurl#1{\textsf{#1}}\fi
\csname PreBibitemsHook\endcsname

\bibitem[\protect\citeauthoryear{Albrecht}{1989}]{albrecht1989aerosols}
\begin{barticle}
\bauthor{\bsnm{Albrecht}, \binits{B.A.}}:
\batitle{Aerosols, cloud microphysics, and fractional cloudiness}.
\bjtitle{Science}
\bvolume{245}(\bissue{4923}),
\bfpage{1227}--\blpage{1230}
(\byear{1989})
\doiurl{10.1126/science.245.4923.1227}
\end{barticle}
\endbibitem

\bibitem[\protect\citeauthoryear{Bellouin et~al.}{2005}]{bellouin2005global}
\begin{barticle}
\bauthor{\bsnm{Bellouin}, \binits{N.}},
\bauthor{\bsnm{Boucher}, \binits{O.}},
\bauthor{\bsnm{Haywood}, \binits{J.}},
\bauthor{\bsnm{Reddy}, \binits{M.S.}}:
\batitle{Global estimate of aerosol direct radiative forcing from satellite measurements}.
\bjtitle{Nature}
\bvolume{438}(\bissue{7071}),
\bfpage{1138}--\blpage{1141}
(\byear{2005})
\end{barticle}
\endbibitem

\bibitem[\protect\citeauthoryear{Battaglia et~al.}{2016}]{battaglia2016interaction}
\begin{botherref}
\oauthor{\bsnm{Battaglia}, \binits{P.}},
\oauthor{\bsnm{Pascanu}, \binits{R.}},
\oauthor{\bsnm{Lai}, \binits{M.}},
\oauthor{\bsnm{Jimenez~Rezende}, \binits{D.}}, et al.:
Interaction networks for learning about objects, relations and physics.
Advances in neural information processing systems
\textbf{29}
(2016)
\end{botherref}
\endbibitem

\bibitem[\protect\citeauthoryear{Chung et~al.}{2005}]{chung2005global}
\begin{botherref}
\oauthor{\bsnm{Chung}, \binits{C.E.}},
\oauthor{\bsnm{Ramanathan}, \binits{V.}},
\oauthor{\bsnm{Kim}, \binits{D.}},
\oauthor{\bsnm{Podgorny}, \binits{I.}}:
Global anthropogenic aerosol direct forcing derived from satellite and ground-based observations.
Journal of Geophysical Research: Atmospheres
\textbf{110}(D24)
(2005)
\end{botherref}
\endbibitem

\bibitem[\protect\citeauthoryear{Fey and Lenssen}{2019}]{fey2019fast}
\begin{botherref}
\oauthor{\bsnm{Fey}, \binits{M.}},
\oauthor{\bsnm{Lenssen}, \binits{J.E.}}:
Fast graph representation learning with pytorch geometric.
arXiv preprint arXiv:1903.02428
(2019)
\end{botherref}
\endbibitem

\bibitem[\protect\citeauthoryear{Fierce et~al.}{2020}]{fierce2020radiative}
\begin{barticle}
\bauthor{\bsnm{Fierce}, \binits{L.}},
\bauthor{\bsnm{Onasch}, \binits{T.B.}},
\bauthor{\bsnm{Cappa}, \binits{C.D.}},
\bauthor{\bsnm{Mazzoleni}, \binits{C.}},
\bauthor{\bsnm{China}, \binits{S.}},
\bauthor{\bsnm{Bhandari}, \binits{J.}},
\bauthor{\bsnm{Davidovits}, \binits{P.}},
\bauthor{\bsnm{Fischer}, \binits{D.A.}},
\bauthor{\bsnm{Helgestad}, \binits{T.}},
\bauthor{\bsnm{Lambe}, \binits{A.T.}}, \betal:
\batitle{Radiative absorption enhancements by black carbon controlled by particle-to-particle heterogeneity in composition}.
\bjtitle{Proceedings of the National Academy of Sciences}
\bvolume{117}(\bissue{10}),
\bfpage{5196}--\blpage{5203}
(\byear{2020})
\end{barticle}
\endbibitem

\bibitem[\protect\citeauthoryear{Fierce et~al.}{2017}]{fierce2017toward}
\begin{barticle}
\bauthor{\bsnm{Fierce}, \binits{L.}},
\bauthor{\bsnm{Riemer}, \binits{N.}},
\bauthor{\bsnm{Bond}, \binits{T.C.}}:
\batitle{Toward reduced representation of mixing state for simulating aerosol effects on climate}.
\bjtitle{Bulletin of the American Meteorological Society}
\bvolume{98}(\bissue{5}),
\bfpage{971}--\blpage{980}
(\byear{2017})
\end{barticle}
\endbibitem

\bibitem[\protect\citeauthoryear{Fierce et~al.}{2024}]{fierce2024quantifying}
\begin{botherref}
\oauthor{\bsnm{Fierce}, \binits{L.}},
\oauthor{\bsnm{Yao}, \binits{Y.}},
\oauthor{\bsnm{Easter}, \binits{R.}},
\oauthor{\bsnm{Ma}, \binits{P.-L.}},
\oauthor{\bsnm{Sun}, \binits{J.}},
\oauthor{\bsnm{Wan}, \binits{H.}},
\oauthor{\bsnm{Zhang}, \binits{K.}}:
Quantifying structural errors in cloud condensation nuclei activity from reduced representation of aerosol size distributions.
Journal of Aerosol Science,
106388
(2024)
\end{botherref}
\endbibitem

\bibitem[\protect\citeauthoryear{Grabowski and Abade}{2017}]{grabowski2017broadening}
\begin{barticle}
\bauthor{\bsnm{Grabowski}, \binits{W.W.}},
\bauthor{\bsnm{Abade}, \binits{G.C.}}:
\batitle{Broadening of cloud droplet spectra through eddy hopping: Turbulent adiabatic parcel simulations}.
\bjtitle{Journal of the Atmospheric Sciences}
\bvolume{74}(\bissue{5}),
\bfpage{1485}--\blpage{1493}
(\byear{2017})
\doiurl{10.1175/JAS-D-17-0043.1}
\end{barticle}
\endbibitem

\bibitem[\protect\citeauthoryear{Gilmer et~al.}{2017}]{gilmer2017neural}
\begin{bchapter}
\bauthor{\bsnm{Gilmer}, \binits{J.}},
\bauthor{\bsnm{Schoenholz}, \binits{S.S.}},
\bauthor{\bsnm{Riley}, \binits{P.F.}},
\bauthor{\bsnm{Vinyals}, \binits{O.}},
\bauthor{\bsnm{Dahl}, \binits{G.E.}}:
\bctitle{Neural message passing for quantum chemistry}.
In: \bbtitle{International Conference on Machine Learning},
pp. \bfpage{1263}--\blpage{1272}
(\byear{2017}).
\bcomment{PMLR}
\end{bchapter}
\endbibitem

\bibitem[\protect\citeauthoryear{Gautam et~al.}{2024}]{gautam2024chemical}
\begin{botherref}
\oauthor{\bsnm{Gautam}, \binits{T.}},
\oauthor{\bsnm{Vandergrift}, \binits{G.W.}},
\oauthor{\bsnm{Lata}, \binits{N.N.}},
\oauthor{\bsnm{Cheng}, \binits{Z.}},
\oauthor{\bsnm{Rahman}, \binits{A.}},
\oauthor{\bsnm{Minke}, \binits{A.}},
\oauthor{\bsnm{Lai}, \binits{Z.}},
\oauthor{\bsnm{Dexheimer}, \binits{D.N.}},
\oauthor{\bsnm{Zhang}, \binits{D.}},
\oauthor{\bsnm{Marcus}, \binits{M.A.}}, et al.:
Chemical insights into the molecular composition of organic aerosols in the urban region of houston, texas.
ACS ES\&T Air
(2024)
\end{botherref}
\endbibitem

\bibitem[\protect\citeauthoryear{Healy et~al.}{2014}]{healy2014single}
\begin{barticle}
\bauthor{\bsnm{Healy}, \binits{R.M.}},
\bauthor{\bsnm{Riemer}, \binits{N.}},
\bauthor{\bsnm{Wenger}, \binits{J.C.}},
\bauthor{\bsnm{Murphy}, \binits{M.}},
\bauthor{\bsnm{West}, \binits{M.}},
\bauthor{\bsnm{Poulain}, \binits{L.}},
\bauthor{\bsnm{Wiedensohler}, \binits{A.}},
\bauthor{\bsnm{O'Connor}, \binits{I.P.}},
\bauthor{\bsnm{McGillicuddy}, \binits{E.}},
\bauthor{\bsnm{Sodeau}, \binits{J.R.}}, \betal:
\batitle{Single particle diversity and mixing state measurements}.
\bjtitle{Atmospheric Chemistry and Physics}
\bvolume{14}(\bissue{12}),
\bfpage{6289}--\blpage{6299}
(\byear{2014})
\end{barticle}
\endbibitem

\bibitem[\protect\citeauthoryear{Kingma and Ba}{2014}]{kingma2014adam}
\begin{botherref}
\oauthor{\bsnm{Kingma}, \binits{D.P.}},
\oauthor{\bsnm{Ba}, \binits{J.}}:
Adam: A method for stochastic optimization.
arXiv preprint arXiv:1412.6980
(2014)
\end{botherref}
\endbibitem

\bibitem[\protect\citeauthoryear{Kumar and Vantassel}{2022}]{kumar2022gns}
\begin{botherref}
\oauthor{\bsnm{Kumar}, \binits{K.}},
\oauthor{\bsnm{Vantassel}, \binits{J.}}:
Gns: A generalizable graph neural network-based simulator for particulate and fluid modeling.
arXiv preprint arXiv:2211.10228
(2022)
\end{botherref}
\endbibitem

\bibitem[\protect\citeauthoryear{Masson-Delmotte et~al.}{2021}]{masson2021climate}
\begin{botherref}
\oauthor{\bsnm{Masson-Delmotte}, \binits{V.}},
\oauthor{\bsnm{Zhai}, \binits{P.}},
\oauthor{\bsnm{Pirani}, \binits{A.}},
\oauthor{\bsnm{Connors}, \binits{S.L.}},
\oauthor{\bsnm{P{\'e}an}, \binits{C.}},
\oauthor{\bsnm{Berger}, \binits{S.}},
\oauthor{\bsnm{Caud}, \binits{N.}},
\oauthor{\bsnm{Chen}, \binits{Y.}},
\oauthor{\bsnm{Goldfarb}, \binits{L.}},
\oauthor{\bsnm{Gomis}, \binits{M.}}, et al.:
Climate change 2021: the physical science basis.
Contribution of working group I to the sixth assessment report of the intergovernmental panel on climate change
\textbf{2}
(2021)
\end{botherref}
\endbibitem

\bibitem[\protect\citeauthoryear{Morrison et~al.}{2020}]{morrison2020confronting}
\begin{barticle}
\bauthor{\bsnm{Morrison}, \binits{H.}},
\bauthor{\bsnm{Lier-Walqui}, \binits{M.}},
\bauthor{\bsnm{Fridlind}, \binits{A.M.}},
\bauthor{\bsnm{Grabowski}, \binits{W.W.}},
\bauthor{\bsnm{Harrington}, \binits{J.Y.}},
\bauthor{\bsnm{Hoose}, \binits{C.}},
\bauthor{\bsnm{Korolev}, \binits{A.}},
\bauthor{\bsnm{Kumjian}, \binits{M.R.}},
\bauthor{\bsnm{Milbrandt}, \binits{J.A.}},
\bauthor{\bsnm{Pawlowska}, \binits{H.}}, \betal:
\batitle{Confronting the challenge of modeling cloud and precipitation microphysics}.
\bjtitle{Journal of advances in modeling earth systems}
\bvolume{12}(\bissue{8}),
\bfpage{2019}--\blpage{001689}
(\byear{2020})
\end{barticle}
\endbibitem

\bibitem[\protect\citeauthoryear{Paszke et~al.}{2019}]{paszke2019pytorch}
\begin{botherref}
\oauthor{\bsnm{Paszke}, \binits{A.}},
\oauthor{\bsnm{Gross}, \binits{S.}},
\oauthor{\bsnm{Massa}, \binits{F.}},
\oauthor{\bsnm{Lerer}, \binits{A.}},
\oauthor{\bsnm{Bradbury}, \binits{J.}},
\oauthor{\bsnm{Chanan}, \binits{G.}},
\oauthor{\bsnm{Killeen}, \binits{T.}},
\oauthor{\bsnm{Lin}, \binits{Z.}},
\oauthor{\bsnm{Gimelshein}, \binits{N.}},
\oauthor{\bsnm{Antiga}, \binits{L.}}, et al.:
Pytorch: An imperative style, high-performance deep learning library.
Advances in neural information processing systems
\textbf{32}
(2019)
\end{botherref}
\endbibitem

\bibitem[\protect\citeauthoryear{Riemer et~al.}{2019}]{riemer2019aerosol}
\begin{barticle}
\bauthor{\bsnm{Riemer}, \binits{N.}},
\bauthor{\bsnm{Ault}, \binits{A.}},
\bauthor{\bsnm{West}, \binits{M.}},
\bauthor{\bsnm{Craig}, \binits{R.}},
\bauthor{\bsnm{Curtis}, \binits{J.}}:
\batitle{Aerosol mixing state: Measurements, modeling, and impacts}.
\bjtitle{Reviews of Geophysics}
\bvolume{57}(\bissue{2}),
\bfpage{187}--\blpage{249}
(\byear{2019})
\end{barticle}
\endbibitem

\bibitem[\protect\citeauthoryear{Rosenfeld et~al.}{2008}]{rosenfeld2008flood}
\begin{barticle}
\bauthor{\bsnm{Rosenfeld}, \binits{D.}},
\bauthor{\bsnm{Lohmann}, \binits{U.}},
\bauthor{\bsnm{Raga}, \binits{G.B.}},
\bauthor{\bsnm{O'Dowd}, \binits{C.D.}},
\bauthor{\bsnm{Kulmala}, \binits{M.}},
\bauthor{\bsnm{Fuzzi}, \binits{S.}},
\bauthor{\bsnm{Reissell}, \binits{A.}},
\bauthor{\bsnm{Andreae}, \binits{M.O.}}:
\batitle{Flood or drought: how do aerosols affect precipitation?}
\bjtitle{science}
\bvolume{321}(\bissue{5894}),
\bfpage{1309}--\blpage{1313}
(\byear{2008})
\doiurl{10.1126/science.1160606}
\end{barticle}
\endbibitem

\bibitem[\protect\citeauthoryear{Riemer et~al.}{2009}]{riemer2009simulating}
\begin{botherref}
\oauthor{\bsnm{Riemer}, \binits{N.}},
\oauthor{\bsnm{West}, \binits{M.}},
\oauthor{\bsnm{Zaveri}, \binits{R.A.}},
\oauthor{\bsnm{Easter}, \binits{R.C.}}:
Simulating the evolution of soot mixing state with a particle-resolved aerosol model.
Journal of Geophysical Research: Atmospheres
\textbf{114}(D9)
(2009)
\end{botherref}
\endbibitem

\bibitem[\protect\citeauthoryear{Sch{\"u}tt et~al.}{2017}]{schutt2017quantum}
\begin{barticle}
\bauthor{\bsnm{Sch{\"u}tt}, \binits{K.T.}},
\bauthor{\bsnm{Arbabzadah}, \binits{F.}},
\bauthor{\bsnm{Chmiela}, \binits{S.}},
\bauthor{\bsnm{M{\"u}ller}, \binits{K.R.}},
\bauthor{\bsnm{Tkatchenko}, \binits{A.}}:
\batitle{Quantum-chemical insights from deep tensor neural networks}.
\bjtitle{Nature communications}
\bvolume{8}(\bissue{1}),
\bfpage{13890}
(\byear{2017})
\end{barticle}
\endbibitem

\bibitem[\protect\citeauthoryear{Sanchez-Gonzalez et~al.}{2020}]{sanchez2020learning}
\begin{bchapter}
\bauthor{\bsnm{Sanchez-Gonzalez}, \binits{A.}},
\bauthor{\bsnm{Godwin}, \binits{J.}},
\bauthor{\bsnm{Pfaff}, \binits{T.}},
\bauthor{\bsnm{Ying}, \binits{R.}},
\bauthor{\bsnm{Leskovec}, \binits{J.}},
\bauthor{\bsnm{Battaglia}, \binits{P.}}:
\bctitle{Learning to simulate complex physics with graph networks}.
In: \bbtitle{International Conference on Machine Learning},
pp. \bfpage{8459}--\blpage{8468}
(\byear{2020}).
\bcomment{PMLR}
\end{bchapter}
\endbibitem

\bibitem[\protect\citeauthoryear{Schwarz et~al.}{2008}]{schwarz2008measurement}
\begin{botherref}
\oauthor{\bsnm{Schwarz}, \binits{J.P.}},
\oauthor{\bsnm{Gao}, \binits{R.}},
\oauthor{\bsnm{Spackman}, \binits{J.}},
\oauthor{\bsnm{Watts}, \binits{L.}},
\oauthor{\bsnm{Thomson}, \binits{D.}},
\oauthor{\bsnm{Fahey}, \binits{D.}},
\oauthor{\bsnm{Ryerson}, \binits{T.}},
\oauthor{\bsnm{Peischl}, \binits{J.}},
\oauthor{\bsnm{Holloway}, \binits{J.}},
\oauthor{\bsnm{Trainer}, \binits{M.}}, et al.:
Measurement of the mixing state, mass, and optical size of individual black carbon particles in urban and biomass burning emissions.
Geophysical Research Letters
\textbf{35}(13)
(2008)
\end{botherref}
\endbibitem

\bibitem[\protect\citeauthoryear{Scarselli et~al.}{2008}]{scarselli2008graph}
\begin{barticle}
\bauthor{\bsnm{Scarselli}, \binits{F.}},
\bauthor{\bsnm{Gori}, \binits{M.}},
\bauthor{\bsnm{Tsoi}, \binits{A.C.}},
\bauthor{\bsnm{Hagenbuchner}, \binits{M.}},
\bauthor{\bsnm{Monfardini}, \binits{G.}}:
\batitle{The graph neural network model}.
\bjtitle{IEEE transactions on neural networks}
\bvolume{20}(\bissue{1}),
\bfpage{61}--\blpage{80}
(\byear{2008})
\end{barticle}
\endbibitem

\bibitem[\protect\citeauthoryear{Shima et~al.}{2009}]{shima2009super}
\begin{barticle}
\bauthor{\bsnm{Shima}, \binits{S.-i.}},
\bauthor{\bsnm{Kusano}, \binits{K.}},
\bauthor{\bsnm{Kawano}, \binits{A.}},
\bauthor{\bsnm{Sugiyama}, \binits{T.}},
\bauthor{\bsnm{Kawahara}, \binits{S.}}:
\batitle{The super-droplet method for the numerical simulation of clouds and precipitation: A particle-based and probabilistic microphysics model coupled with a non-hydrostatic model}.
\bjtitle{Quarterly Journal of the Royal Meteorological Society: A journal of the atmospheric sciences, applied meteorology and physical oceanography}
\bvolume{135}(\bissue{642}),
\bfpage{1307}--\blpage{1320}
(\byear{2009})
\end{barticle}
\endbibitem

\bibitem[\protect\citeauthoryear{Stier et~al.}{2013}]{stier2013host}
\begin{barticle}
\bauthor{\bsnm{Stier}, \binits{P.}},
\bauthor{\bsnm{Schutgens}, \binits{N.A.}},
\bauthor{\bsnm{Bellouin}, \binits{N.}},
\bauthor{\bsnm{Bian}, \binits{H.}},
\bauthor{\bsnm{Boucher}, \binits{O.}},
\bauthor{\bsnm{Chin}, \binits{M.}},
\bauthor{\bsnm{Ghan}, \binits{S.}},
\bauthor{\bsnm{Huneeus}, \binits{N.}},
\bauthor{\bsnm{Kinne}, \binits{S.}},
\bauthor{\bsnm{Lin}, \binits{G.}}, \betal:
\batitle{Host model uncertainties in aerosol radiative forcing estimates: results from the aerocom prescribed intercomparison study}.
\bjtitle{Atmospheric Chemistry and Physics}
\bvolume{13}(\bissue{6}),
\bfpage{3245}--\blpage{3270}
(\byear{2013})
\end{barticle}
\endbibitem

\bibitem[\protect\citeauthoryear{Twomey}{1977}]{twomey1977influence}
\begin{barticle}
\bauthor{\bsnm{Twomey}, \binits{S.}}:
\batitle{The influence of pollution on the shortwave albedo of clouds}.
\bjtitle{Journal of the atmospheric sciences}
\bvolume{34}(\bissue{7}),
\bfpage{1149}--\blpage{1152}
(\byear{1977})
\doiurl{10.1175/1520-0469(1977)034<1149:TIOPOT>2.0.CO;2}
\end{barticle}
\endbibitem

\bibitem[\protect\citeauthoryear{Veli{\v{c}}kovi{\'c} et~al.}{2017}]{velivckovic2017graph}
\begin{botherref}
\oauthor{\bsnm{Veli{\v{c}}kovi{\'c}}, \binits{P.}},
\oauthor{\bsnm{Cucurull}, \binits{G.}},
\oauthor{\bsnm{Casanova}, \binits{A.}},
\oauthor{\bsnm{Romero}, \binits{A.}},
\oauthor{\bsnm{Lio}, \binits{P.}},
\oauthor{\bsnm{Bengio}, \binits{Y.}}:
Graph attention networks.
arXiv preprint arXiv:1710.10903
(2017)
\end{botherref}
\endbibitem

\bibitem[\protect\citeauthoryear{Wu et~al.}{2020}]{wu2020comprehensive}
\begin{barticle}
\bauthor{\bsnm{Wu}, \binits{Z.}},
\bauthor{\bsnm{Pan}, \binits{S.}},
\bauthor{\bsnm{Chen}, \binits{F.}},
\bauthor{\bsnm{Long}, \binits{G.}},
\bauthor{\bsnm{Zhang}, \binits{C.}},
\bauthor{\bsnm{Yu}, \binits{P.S.}}:
\batitle{A comprehensive survey on graph neural networks}.
\bjtitle{IEEE Transactions on Neural Networks and Learning Systems}
\bvolume{32}(\bissue{1}),
\bfpage{4}--\blpage{24}
(\byear{2020})
\end{barticle}
\endbibitem

\bibitem[\protect\citeauthoryear{Zaveri et~al.}{2008}]{zaveri2008development}
\begin{botherref}
\oauthor{\bsnm{Zaveri}, \binits{R.A.}}, et al.:
Development and evaluation of an explicit aqueous-phase chemistry module in the aerosol dynamics, gas- and aerosol-phase chemistry, and transport model, partmc-mosaic.
Journal of Geophysical Research: Atmospheres
\textbf{113}(D7)
(2008)
\end{botherref}
\endbibitem

\bibitem[\protect\citeauthoryear{Zheng et~al.}{2021}]{zheng2021quantifying}
\begin{barticle}
\bauthor{\bsnm{Zheng}, \binits{Z.}},
\bauthor{\bsnm{West}, \binits{M.}},
\bauthor{\bsnm{Zhao}, \binits{L.}},
\bauthor{\bsnm{Ma}, \binits{P.-L.}},
\bauthor{\bsnm{Liu}, \binits{X.}},
\bauthor{\bsnm{Riemer}, \binits{N.}}:
\batitle{Quantifying the structural uncertainty of the aerosol mixing state representation in a modal model}.
\bjtitle{Atmospheric Chemistry and Physics}
\bvolume{21}(\bissue{23}),
\bfpage{17727}--\blpage{17741}
(\byear{2021})
\end{barticle}
\endbibitem

\end{thebibliography}

\end{document}